\documentclass[main]{aa}  
\usepackage{graphicx}
\usepackage{txfonts}
\usepackage{amsmath}
\usepackage[utf8]{inputenc}
\usepackage{xcolor}
\usepackage{verbatim}

\bibpunct{(}{)}{;}{a}{}{,} 

\def\apjl{{ApJL}}
\def\apjs{{ApJS}}

\def\farcs{\hbox{$.\mkern-4mu^{\prime\prime}$}}

\def\hal{H$\alpha$}
\def\hb{H$\beta$}

\def\redd{$\lambda_{\rm Edd}$}
\def\lax{{$\mathrel{\hbox{\rlap{\hbox{\lower4pt\hbox{$\sim$}}}\hbox{$<$}}}$}}
\def\gax{{$\mathrel{\hbox{\rlap{\hbox{\lower4pt\hbox{$\sim$}}}\hbox{$>$}}}$}}
\def\simlt{\lower.5ex\hbox{$\; \buildrel < \over \sim \;$}}
\def\simgt{\lower.5ex\hbox{$\; \buildrel > \over \sim \;$}}

\def\mbh{{$M_{\rm BH}$}}
\def\micron{{$\mu$m}}

\def\cm2{cm$^{-2}$}

\def\oiii{[\ion{O}{III}]}

\def\loiii{$L_{\rm [O\ III]}$}
\def\lbol{$L_{{\rm bol}}$}
\def\ledd{$L_{{\rm Edd}}$}

\def\l5100{$L_{5100}$}
\def\ll5100{$\log L_{\rm 5100}$}

\def\-->{$\rightarrow$}
\def\sf200{${\rm SF}_{\rm 200 days}$}

\begin{document}

\title{Asymmetric torus variability in active galactic nuclei driven by global brightening and dimming
}

\author{Suyeon Son\inst{1,2} \and Minjin Kim\inst{3,2} \and Luis C. Ho\inst{1,4}}

\institute{Kavli Institute for Astronomy and Astrophysics, Peking University, Beijing 100871, China\\
\email{sson.astro@gmail.com, mkim.astro@yonsei.ac.kr}
\and
Department of Astronomy and Atmospheric Sciences,
Kyungpook National University, Daegu 41566, Korea
\and
Department of Astronomy, Yonsei University, Seoul 03722, Korea
\and
Department of Astronomy, School of Physics, Peking University, Beijing 100871, China
}

\date{Received}

\abstract{
Temporal asymmetry in the flux variability of active galactic nuclei (AGNs) offers key insights into the physical mechanisms driving AGN variability. In this study, we investigated the variability of the torus by analyzing temporal asymmetry in the mid-infrared (MIR) continuum. We compared ensemble structure functions between the brightening and dimming phases for AGNs at $0.15<z<0.4$, using monitoring data in the optical from the Zwicky Transient Facility and in the MIR from the Near-Earth Object Wide-field Infrared Survey Explorer. We found that AGNs with bluer optical-to-MIR colors exhibit positive temporal asymmetry in the MIR, indicating that their variability amplitude is larger when brightening. Conversely, those with redder colors show negative asymmetry, exhibiting larger variability amplitude when decaying. However, there is no significant temporal asymmetry in the $g$-band variability driven by the accretion disk, suggesting that the temporal asymmetry in the MIR continuum primarily originates from intrinsic processes in the torus instead of the reflection of the ultraviolet-optical variability from the accretion disk. Analysis of the composite light curves revealed that AGNs with bluer optical-to-MIR colors tend to brighten gradually in the MIR, leading to the observed temporal asymmetry. This finding suggests that hot-dust-rich AGNs evolve with a gradual decline in hot dust emission, while hot-dust-poor AGNs are associated with a steady increase.
}


\titlerunning{SF Asymmetry}
\authorrunning{Son et al.}

\maketitle

\section{Introduction}

Active galactic nuclei (AGNs) exhibit flux variability across all observed wavelengths on timescales ranging from days to years. This variability is commonly characterized using stochastic Gaussian processes. Specifically, the ultraviolet and optical continuum light curves, originating from the accretion disk (AD), are often described using the damped random walk (DRW) model. In this framework, the variability amplitude increases with timescale below a characteristic threshold, beyond which it asymptotically approaches a constant value (e.g., \citealt{kelly_2009,kozlowski_2010b,macleod_2010}). Although the DRW model provides merely a statistical framework, the variability features it characterizes can offer meaningful constraints on the underlying physical mechanisms. Recent studies suggest that AGN variability marginally deviates from the DRW process, and several alternative models have been introduced (e.g., \citealt{kelly_2014,kasliwal_2017,yu_2025}). A fundamental assumption of these models is that flux variability is symmetric over time, meaning that the amplitudes of brightening (increasing flux) and dimming (decreasing flux) events are identical at a given timescale. However, the extent to which this assumption holds for AGN variability remains an open question. The presence of temporal asymmetry, if confirmed, could provide valuable insights into the driving mechanisms of AGN variability \citep{kawaguchi_1998,hawkins_2002}.

Temporal asymmetry in the flux variability can arise from differences between the timescales of the brightening and dimming phases. For instance, rapid brightening followed by gradual dimming, as seen in tidal disruption events and supernovae, leads to brightening dominating (i.e., positive asymmetry) at short timescales and reverses to dimming dominating (i.e., negative asymmetry) at longer timescales (Fig. 1). Conversely, gradual brightening followed by rapid dimming results in negative asymmetry at short timescales and positive asymmetry at longer timescales. Based on this concept, \citet{kawaguchi_1998} proposed that variability in the ultraviolet-optical continuum from ADs is expected to exhibit positive asymmetry if caused by starbursts and negative asymmetry if caused by disk instabilities on timescales shorter than a few hundred days. Additionally, symmetric variability is expected if the continuum variability originates from gravitational microlensing \citep{hawkins_2002}. Therefore, the signature and degree of the asymmetry in the variability can provide useful information about the physical origin of the variability.

Motivated by the work of \citet{kawaguchi_1998}, previous studies have analyzed temporal asymmetry by comparing the structure functions (SFs) for the brightening (${\rm SF}_+$) and dimming (${\rm SF}_-$) phase, but no consensus has been reached. \cite{hawkins_2002} showed that moderately luminous AGNs (Seyfert galaxies) exhibit negative asymmetry on timescales of $\sim 10-70$ days, while luminous AGNs (quasars) have symmetric variability over timescales of $\sim 1-20$ years. Symmetric variability of quasars was also reported by \cite{bauer_2009} on timescales from weeks to years. On the other hand, using multi-epoch photometric data of Stripe 82 from the Sloan Digital Sky Survey (SDSS), \cite{voevodkin_2011} identified negative asymmetry on timescales of $300-1600$ days. On the contrary, \cite{devries_2005} presented positive asymmetry at $\sim 1-10$ years for quasars. \cite{tachibana_2020}, using machine learning techniques, showed that temporal asymmetry is positive at $\lesssim$ 100 days and turns over to negative at $\gtrsim$ 200 days. They also argued that the degree of asymmetry correlates inversely with the optical luminosity and black hole (BH) mass of the AGN.

\begin{figure}[tp]
\centering
\includegraphics[width=0.5\textwidth]{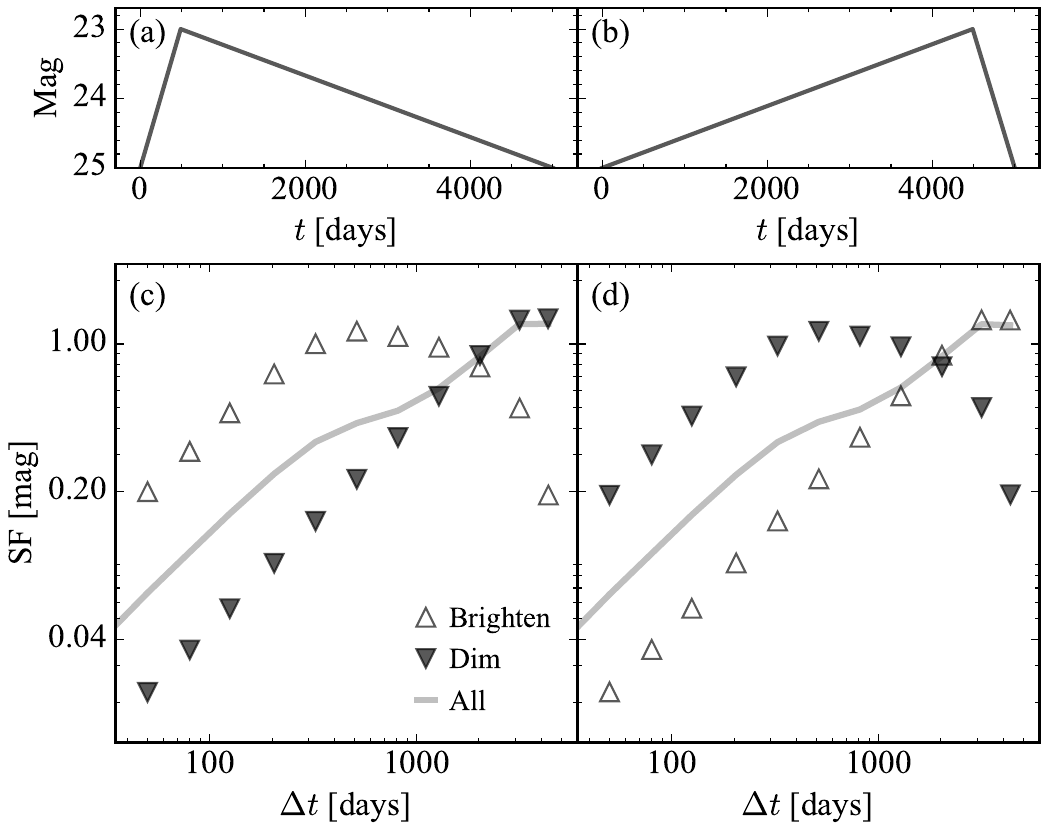}
\caption{Conceptual illustration of structure function (SF) asymmetry. Panel (a) shows an example light curve with a fast rise and slower decay, while panel (b) shows the opposite case with a slow rise and fast decay. The corresponding SFs are decomposed into the total (solid line), brightening (${\rm SF_+}$; up triangle), and fading (${\rm SF_-}$; down triangle) components in panels (c) and (d).
}
\end{figure}

\begin{figure*}[htp!]
\centering
\includegraphics[width=0.32\textwidth]{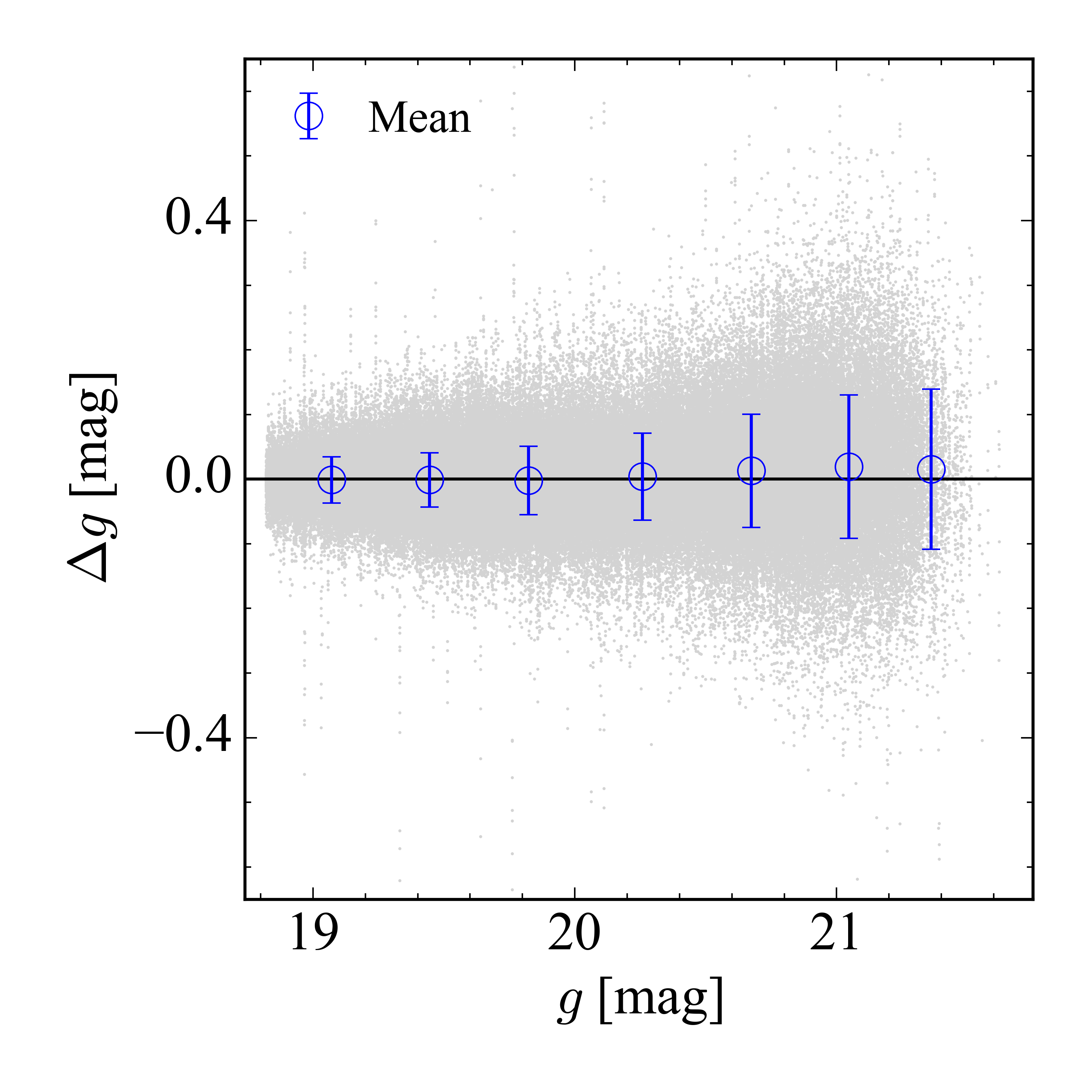}
\includegraphics[width=0.32\textwidth]{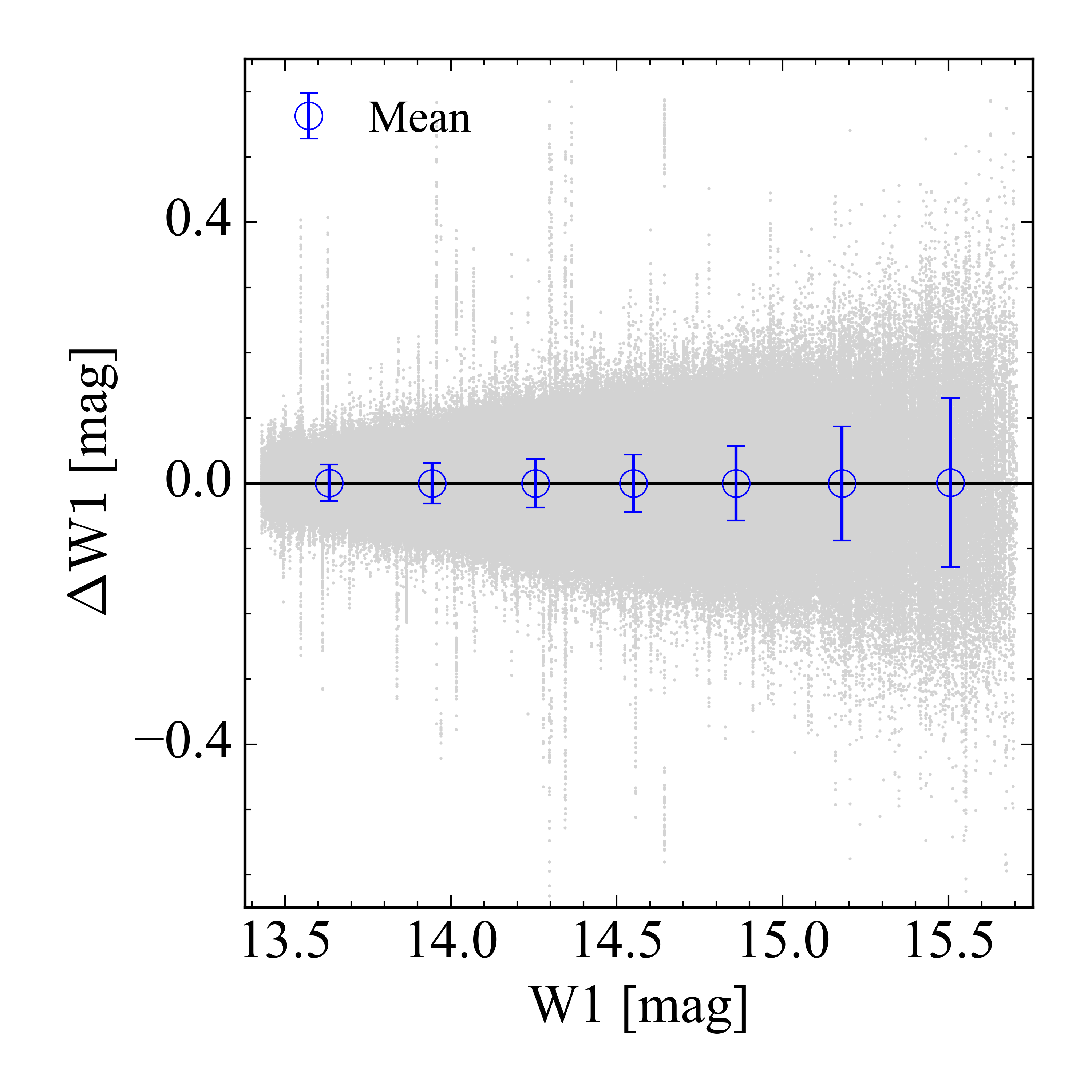}
\includegraphics[width=0.32\textwidth]{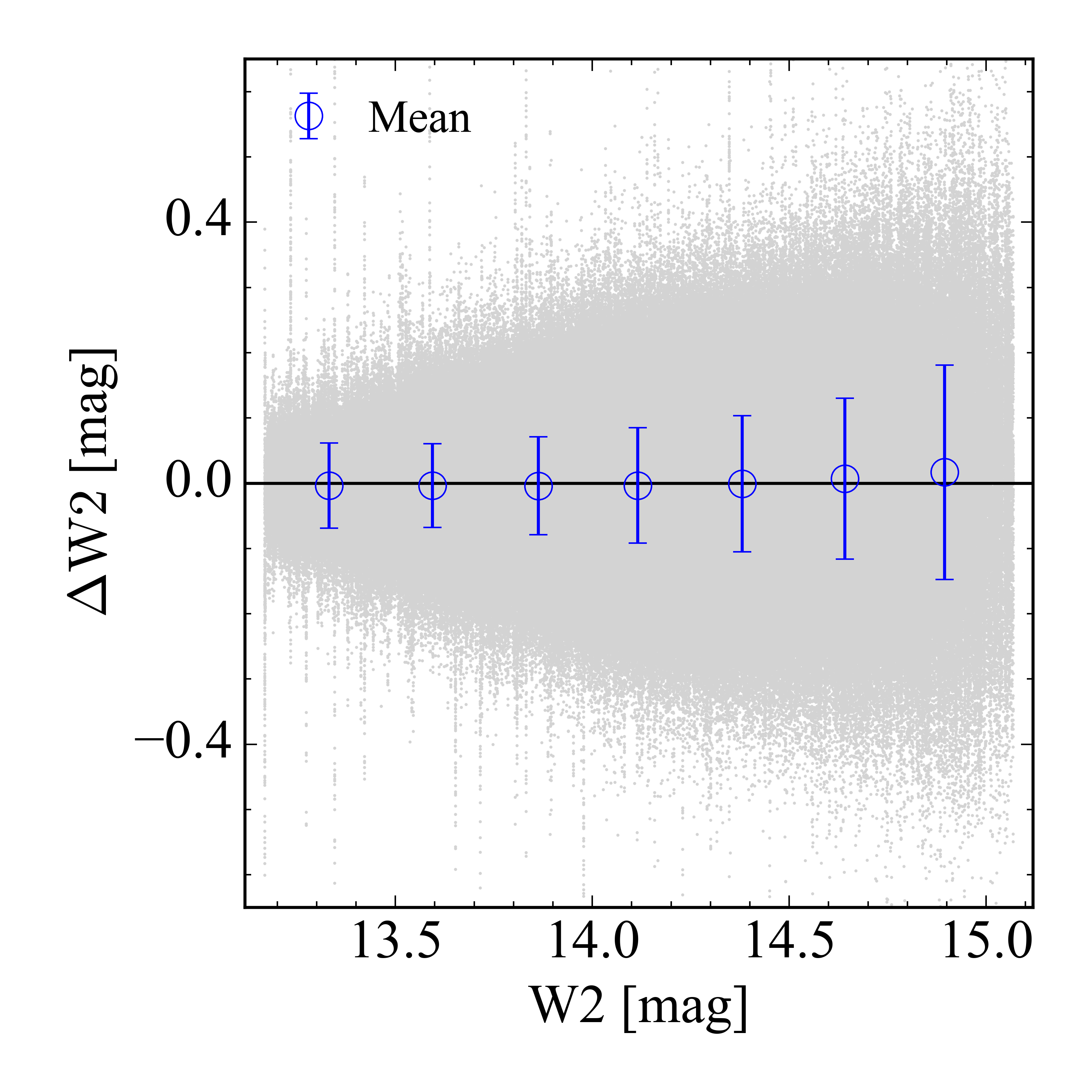}
\caption{Differences between magnitude pairs for inactive galaxies in the $g$ (left), W1 (middle), and W2 (right) bands. Gray points represent individual magnitude differences. Blue open circles denote the mean values of the magnitude differences with 1$\sigma$ error bars.
}
\end{figure*}

The observed positive and negative temporal asymmetry can also arise from global brightening and dimming trends, respectively. While most of the previous studies concluded that there is no clear signature of a global trend, \cite{caplar_2020} argued, based on a comparison between SDSS and HSC photometry, that AGNs dim on average over $14.85$ years. However, this dimming trend can also be an artifact of Malmquist-like bias (e.g., \citealt{macleod_2012}) or filter differences (e.g., \citealt{morganson_2014}). Careful treatment of the systematics must be accounted for when combining data from multiple photometric surveys. Additionally, the majority of previous studies focused on the variability of the ultraviolet-optical continuum. Multiwavelength studies offer a unique opportunity for a more comprehensive exploration of AGN variability and their central structure. 

In this paper, we investigated the temporal asymmetry of the mid-infrared (MIR) and optical continuum in AGNs to explore the underlying mechanism driving torus variability. To examine long-term variability free from unpredictable biases, we analyzed flux variability using homogeneous photometry from 9.1 years of MIR and 5.7 years of optical light curves obtained from the Wide-field Infrared Survey Explorer (WISE; \citealt{wright_2010}) and the Zwicky Transient Facility (ZTF; \citealt{bellm_2019}), respectively. Comparing ensemble SFs for the brightening and dimming phases enables us to probe the temporal asymmetry of both the primary AD emission and reprocessed torus emission. Using the sample and data described in Section 2, we computed ensemble SFs in Section 3. Section 4 presents the results of ensemble SF analyses for various subsamples. The possible origin of the asymmetry is discussed in Section 5, followed by the conclusions in Section 6.

\section{Sample and data}

The initial sample of type 1 AGNs was drawn from the SDSS Data Release 14 quasar catalog \citep{paris_2018}. For type 2 sources, the initial sample consists of galaxies classified as AGNs based on the Baldwin, Phillips, and Terlevich (BPT) diagram \citep{baldwin_1981,veilleux_1987} in the SDSS Data Release 8 of the Max Planck Institute for Astrophysics-Johns Hopkins University (MPA-JHU) catalog \citep{brinchmann_2004}. We selected 7,443 type 1 and 6,855 type 2 AGNs at $0.15<z<0.4$, a redshift range optimized for optical and MIR SF analysis \citep{son_2023b,kim_2024}. The lower redshift limit ($z>0.15$) was adopted to avoid extended sources that potentially introduce systematics in the photometry between various photometric surveys. Since the \hal\ region must be covered by the SDSS spectra for BPT classification, we set an upper redshift limit of $z<0.4$. The narrow range of the redshift also ensures a comparable rest-frame wavelength range corresponding to the W1 and W2 bands, allowing us to trace hot dust components of the torus consistently across the sample.

We estimated the BH mass (\mbh) of type 1 AGNs using the single-epoch virial mass estimator from \cite{ho_2015}, which is based on the FWHM of broad \hb\ (FWHM$_{{\rm H}\beta}$) and the monochromatic luminosity at 5100 \AA\ ($L_{5100}$), given by $\log\ (M_{\rm BH}/M_\odot) = 6.91 + 2\log\ ({\rm FWHM_{{\rm H}\beta}}/1000\ {\rm km\ s^{-1}})+0.533\log\ (L_{5100}/10^{44}\ {\rm erg\ s^{-1}})$. For type 2 AGNs, we adopt the relation between \mbh\ and total stellar mass ($M_*$) from \cite{greene_2020}: $\log\ (M_{\rm BH}/M_\odot) = 7.43 + 1.61\log\ (M_*/3\times10^{10}\,M_\odot)$, with $M_*$ calculated by \cite{kauffmann_2003} from fits to the photometric data as provided in the MPA–JHU catalog.

We derived the bolometric luminosity assuming a bolometric correction based on the observed (not corrected for dust reddening) luminosity of the \oiii\ $\lambda 5007$ line, \lbol\ = 3500\loiii\ \citep{heckman_2004}, a method that can be applied to both AGN types \citep{heckman_2004}. While the \oiii\ luminosity can be affected by reddening \citep{kong_2018}, among type 1 AGNs, the observed \loiii\ correlates more tightly with $L_{5100}$ than the extinction-corrected \loiii\ (\citealt{son_2023b}). We therefore use the \oiii\ luminosity uncorrected for dust extinction. All the spectral measurements (e.g., FWHM$_{{\rm H}\beta}$, $L_{5100}$, and \loiii) from \cite{rakshit_2020} and the MPA-JHU catalog (\citealt{kauffmann_2003, brinchmann_2004, tremonti_2004}) were adapted to derive AGN properties. The Eddington ratio is defined as \redd\ $\equiv$ \lbol/\ledd, where the Eddington luminosity \ledd\ = $1.26 \times 10^{38}$(\mbh/$M_\odot$) erg s$^{-1}$.

To analyze temporal asymmetry in the MIR continuum dominated by the hot dust component of the torus, we utilized ten years of monitoring data taken in the W1 (3.4 \micron) and W2 (4.6 \micron) bands by the Near-Earth Object Wide-field Infrared Survey Explorer (NEOWISE; \citealt{mainzer_2011}). For type 1 AGNs, we additionally examined the asymmetry of the optical continuum variability by utilizing approximately six years of $g$-band monitoring data from ZTF. A matching radius of $2''$ was used to cross-correlate the initial sample with NEOWISE and ZTF data. As the variability analysis is sensitive to outliers, we utilized flagging and binning to minimize this effect. We only use photometric data with \texttt{cc\_flags} $=0$, \texttt{qual\_frame} $> 0$, \texttt{qi\_fact} $>0$, \texttt{saa\_sep} $>0$, and \texttt{moon\_masked} $=0$ for NEOWISE \citep{son_2022} and flag = 0, separation angle $< 0\farcs4$, and $g < 20.2$ mag for ZTF \citep{kim_2024}. Then, we binned the NEOWISE light curves into cadences of $\sim 6$ months following \citet{son_2022}. Unlike the WISE mission, ZTF light curves are irregularly sampled due to the visibility and weather conditions. Therefore, we binned ZTF light curves with the number of epochs $N_{\rm epoch}\geq20$ using mean-shift clustering with a 100-day bandwidth flat kernel with the help of the \texttt{MeanShift} library in the Python package \texttt{scikit-learn}\footnote{\url{https://scikit-learn.org/stable/modules/generated/sklearn.cluster.MeanShift.html}}. 

During the binning process, we performed 3-sigma clipping within each bin for both NEOWISE and ZTF data to remove outliers. The above qualifications resulted in light curves for the W1 and W2 bands of 7,108 type 1 AGNs and $g$-band light curves for 6,585 type 1 AGNs. Additionally, the W1 and W2 band light curves were available for 6,703 type 2 AGNs. Note that the binned data of all final samples have $N_{\rm epoch} \geq 2$. The type 1 AGNs have NEOWISE and ZTF light curves of average length $3655$ and 1947 days, cadence $198$ and 350 days, and $N_{\rm epoch} = 20.6$ and 6.8, respectively. For type 2 AGNs, the NEOWISE light curves have an average length of 3679 days, a cadence of 191 days, and $N_{\rm epoch} = 20.9$.

\subsection{Zero point correction for NEOWISE}

For robust estimation of the temporal asymmetry in the variability, it is crucial to minimize potential systematics in the photometry. During the NEOWISE survey, the focal plane temperature estimated from the beamsplitter mount gradually increased and oscillated over time, mostly due to the orbital decay of the payload. This reduced sensitivity and potentially led to inaccuracies in the flux estimation.\footnote{\url{https://wise2.ipac.caltech.edu/docs/release/neowise/expsup/sec4_2d.html}} As this systematic effect can introduce artificial temporal asymmetry, we performed a photometric correction based on the focal plane temperature at the time of observation. 

To quantify the photometric correction, we began by examining the NEOWISE light curves of inactive galaxies, assumed to be intrinsically non-variable. We randomly selected 20,000 inactive galaxies at $0.15<z<0.4$ labeled as ``unclassifiable'' in the MPA-JHU catalog, indicating that their emission lines are very weak or absent, with no optical signature of AGNs. The light curves of inactive galaxies were processed in the same manner as the AGN sample. A visual inspection of the mean light curve of the inactive galaxies revealed an oscillating pattern and abnormal jumps at MJD > 60400, suggesting the necessity of additional correction. Another photometric correction was therefore applied to the remaining NEOWISE data by aligning it to the AllWISE data, after which we discarded all data observed at MJD > 60400. For that purpose, we first computed average light curves of inactive galaxies with 300-day bins. Second, we adopted the mean flux during AllWISE as the reference point, assuming that the flux of inactive galaxies remains constant, and then calculated zero point offsets for NEOWISE as a function of MJD. The zero point corrections range from 0.007 to 0.036 mag, indicating that their impact on the SF estimation is marginal. Finally, we applied the zero point correction to the NEOWISE light curves of the AGN sample. We note that the AllWISE data were used only for photometric correction, and not for the analysis in this study, due to systematic offsets in the AllWISE light curves observed in inactive galaxies.

\begin{figure}[tp]
\centering
\includegraphics[width=0.5\textwidth]{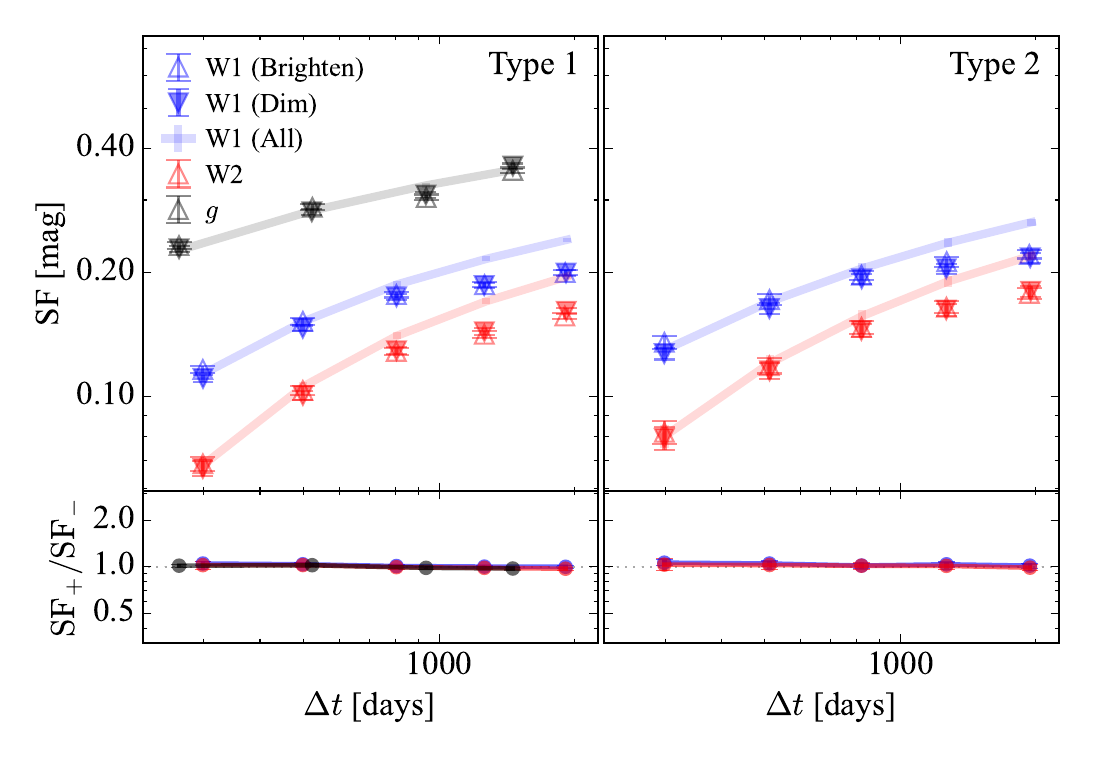}
\caption{Ensemble SFs for type 1 (left) and type 2 (right) AGNs. Open upward and filled downward triangles denote the ensemble SFs for the brightening and dimming phases, respectively. Ensemble SFs for all phases are shown as solid lines. Blue, red, and black symbols correspond to the W1, W2, and $g$ bands, respectively. The ratio of SF$_{+}$ to SF$_{-}$ is shown in the bottom panels.
}
\end{figure}

\begin{figure*}[htp!]
\centering
\includegraphics[width=0.9\textwidth]{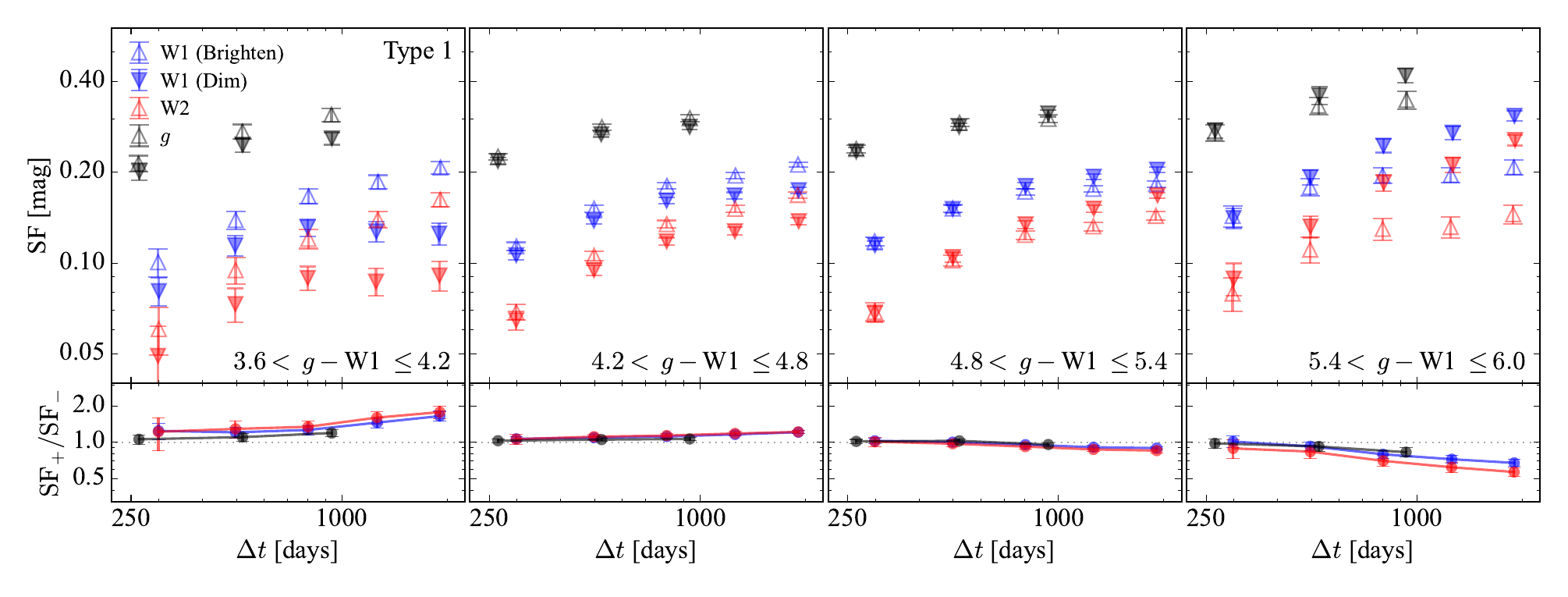}
\includegraphics[width=0.9\textwidth]{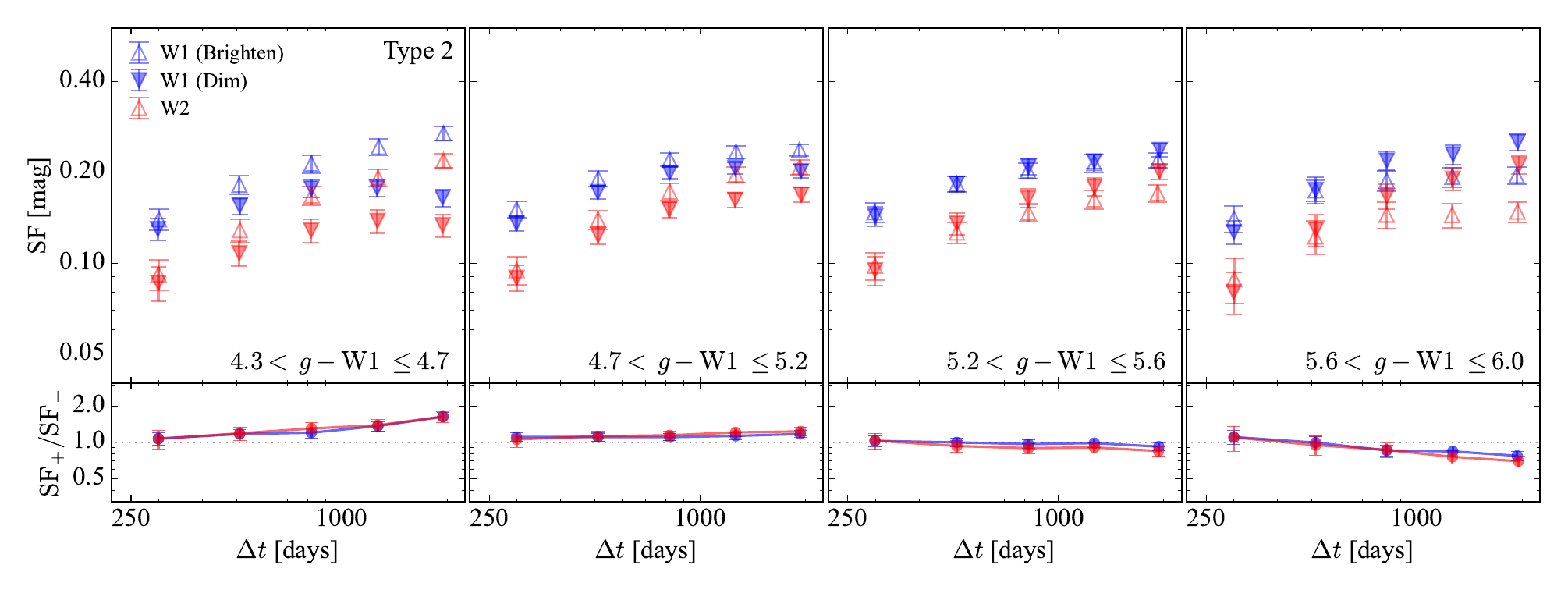}
\caption{Same as Figure 3, but the subsamples are divided by $g-$W1 color for type 1 (top) and type 2 (bottom) AGNs.
}
\end{figure*}

\subsection{Photometric uncertainty}

We did not adopt the flux uncertainties originally provided by each survey, as our analysis relies on binned light curves. Instead, we estimated the flux uncertainties using the sample of inactive galaxies described in Section 2.1. The temperature correction and the zero point correction between the AllWISE and NEOWISE datasets were applied to the NEOWISE light curves following the same procedures used for the sample of AGNs. To quantify the uncertainties in each epoch, we computed the standard deviation of the magnitude differences for all possible pairs within each magnitude bin of the inactive galaxy sample (Fig. 2). The resulting standard deviation, divided by $\sqrt{2}$, was then used to represent the magnitude uncertainty as a function of magnitude.

\subsection{Host galaxy and AD contamination}

This study examines the variability of the torus continuum using the W1 and W2 bands and of the AD continuum using the $g$-band. To robustly isolate these components, it is essential to subtract contributions from the host galaxy and AD, as the W1 and W2 bands can be affected by both, while the $g$-band is primarily contaminated by host emission. However, \cite{son_2023b} showed that for low-$z$ AGNs, such as those in this study, the contribution of the AD to the W1 and W2 bands is negligible. Thus, its effect is disregarded. 

Meanwhile, the contributions from the host emission to the $g$, W1, and W2 bands were estimated through the spectral energy distribution (SED) decomposition. To this end, we constructed SEDs spanning from the optical to the MIR using SDSS ($ugriz$), AllWISE (W1--W4) photometry, and, when available, near-infrared ($JHK_s$) photometry from the Two Micron All Sky Survey (2MASS; \citealt{skrutskie_2006}). We performed SED fitting with semi-empirical templates using the {\tt LePhare} code (\citealt{arnouts_1999, ilbert_2006}). The templates were generated by combining two components: three AGN templates based on the dust properties from \cite{lyu_2017a}, and seven host galaxy templates representing different morphologies from \cite{lyu_2018} and \cite{polletta_2007}; see \citealt{son_2023} for details. The host fluxes in the $g$, W1, and W2 bands, as measured from the best-fit SEDs, are subtracted from the multi-epoch data to remove host contamination. On average, the host contributions of type 1 AGNs in the $g$, W1, and W2 bands are 40\%, 30\%, and 19\%, respectively, suggesting that the correction has a moderate impact on SF estimation.

\begin{figure*}[htp!]
\centering
\includegraphics[width=0.9\textwidth]{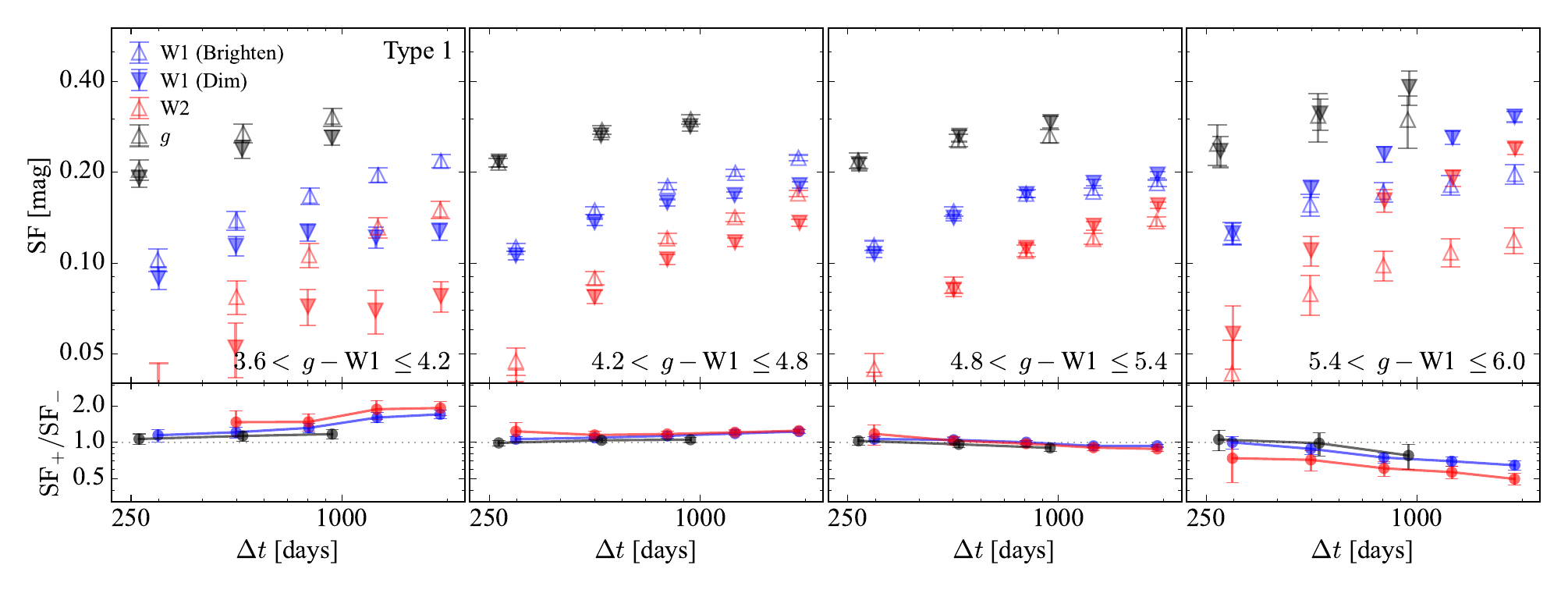}
\includegraphics[width=0.9\textwidth]{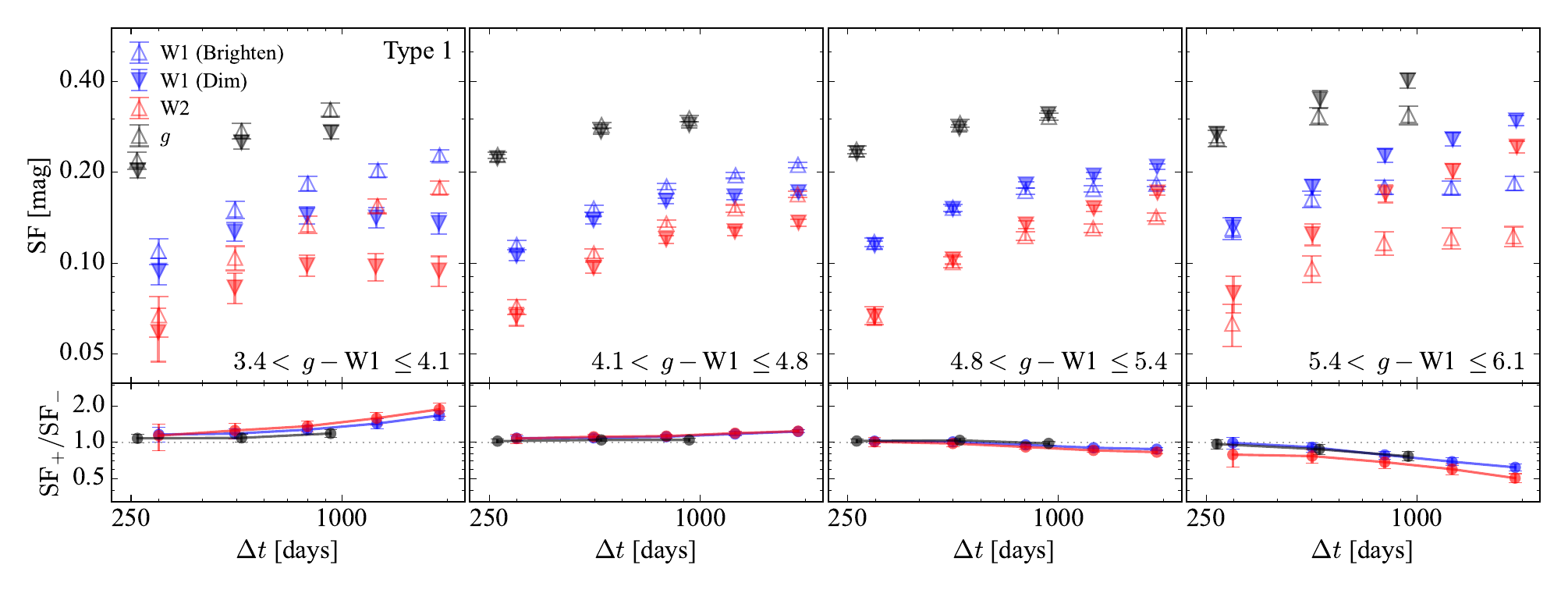}
\includegraphics[width=0.9\textwidth]{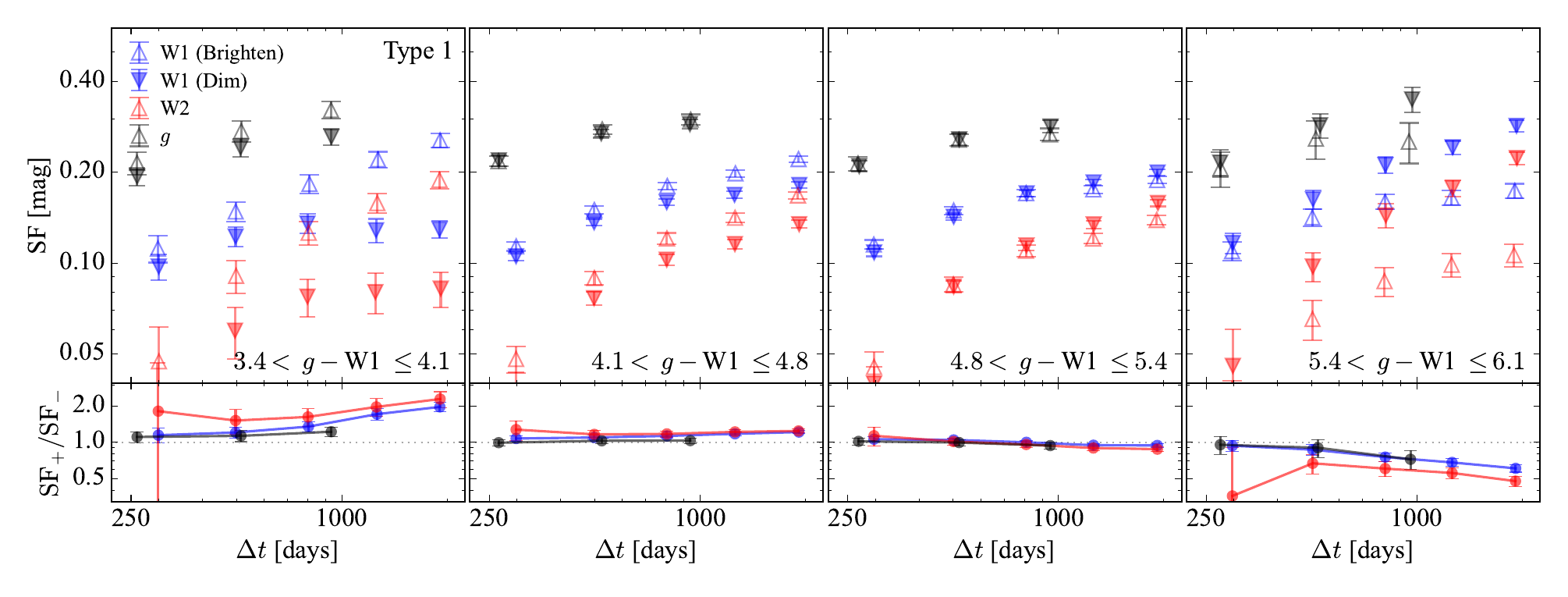}
\caption{Same as Figure 4, but the ensemble SFs are computed by applying brighter magnitude cuts to minimize Eddington bias (top), using subsamples divided by host-subtracted $g-$W1 color (middle), and applying magnitude cuts while using the host-subtracted color (bottom).
}
\end{figure*}

\section{Ensemble SFs}

Studying temporal asymmetry in an individual object is challenging, as the light curve of a single target can be systematically biased due to limited epochs and baseline, and can be easily affected by episodic brightening or dimming. Episodic, long-term trends must be removed to assess reliably the SF asymmetry for individual sources (e.g., \citealt{chen_2015}). To minimize such biases from individual objects, we utilized the ensemble SF, which is defined as the average of SFs from multiple targets at a given time lag. Following the definition of \cite{press_1992}, the ensemble SF is computed using the following form, which is known to reproduce accurately the true SF \citep{kozlowski_2016b,son_2023b,kim_2024}. 
\begin{eqnarray}
    {\rm SF}^2(\Delta t) = \frac{1}{N_{\Delta t, {\rm pair}}} \sum_{i=1}^{N_{\Delta t, {\rm pair}}}
    (m(t) - m(t+\Delta t))^2 \nonumber \\
    - \sigma_m^2(t) - \sigma_m^2(t+\Delta t), 
\end{eqnarray}
\noindent
where $N_{\Delta t, {\rm pair}}$ is the number of observed magnitude pairs with time lag $\Delta t$, $m$ is the observed magnitude, and $\sigma_m$ is the uncertainty of the magnitude. To prevent sources with long baselines from dominating the SF at large $\Delta t$, we only used sufficiently long light curves, because the SF at large $\Delta t$ is predominantly computed from the objects with long light curves. For this reason, we selected only targets with more than 18 and 5 epochs in the WISE and ZTF datasets, respectively. To exclude AGNs heavily contaminated by host galaxy emission, which can lead to inaccurate photometry of the AGN component, we included only targets with host contributions less than 0.5 in the light curves,

\begin{figure}[htp!]
\centering
\includegraphics[width=0.5\textwidth]{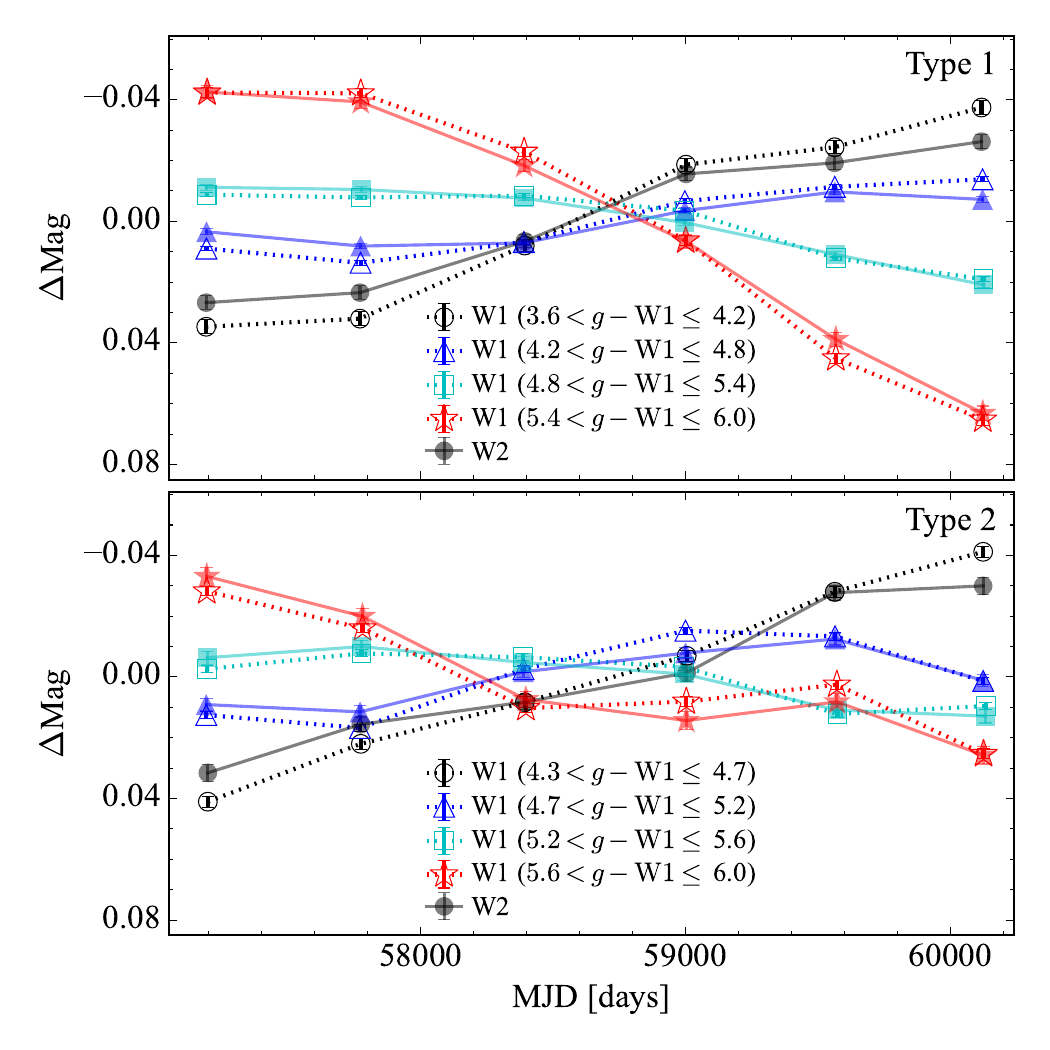}
\caption{Average composite light curves for type 1 (top) and type 2 (bottom) AGNs, grouped by their $g-$W1 color.
}
\end{figure}
\noindent in both the W1 and $g$ bands for type 1 AGNs and in the W1 band for type 2 AGNs. Finally, ensemble SFs were computed for 3558 W1 and W2 band light curves and 3611 $g-$band light curves for type 1 AGNs, with average lengths of 3303 and 2073 days, cadence of 182 and 336 days, and $N_{\rm epoch} = 19.1$ and 7.2, respectively. For type 2 AGNs, 1122 W1 and W2 band light curves were used, with an average length of 3310 days, a cadence of 182 days, and $N_{\rm epoch} = 19.1$.

\section{Results}
We examined the temporal asymmetry in AGN continuum by comparing ensemble SFs between brightening and dimming phases. Following \cite{kawaguchi_1998}, the SF for the brightening phase (SF$_+$) is calculated using photometric pairs in which the later observation is brighter, while SF for the dimming phase (SF$_-$) is derived using magnitude pairs where the later observation is fainter. At a given timescale, the SF is expected to exhibit positive asymmetry (SF$_+ >$ SF$_-$) if brightening is more dominant, and negative asymmetry (SF$_- >$ SF$_+$) if dimming is more dominant. Figure 3 shows ensemble SFs for the brightening and dimming phases of type 1 and type 2 AGNs. The SFs flatten at longer timescales in all bands, and the variability amplitude is larger at shorter wavelengths, consistent with previous studies \cite[e.g.,][]{devries_2003,ivezic_2004,vandenberk_2004,kim_2024}. The torus variability in the W1 and W2 bands appears to be temporally symmetric for both type 1 and type 2 AGNs, regardless of the timescale. For type 1 AGNs, the $g$-band variability from the AD also shows no temporal asymmetry. 

We examined whether SF asymmetry arises in type 1 AGN subsamples divided by fundamental properties of AGNs. We ensured the statistical robustness of the SF asymmetry analysis by requiring a minimum of 200 sources in each subsample. There is no significant SF asymmetry in MIR and optical continuum variability across the subsamples divided by AGN properties, 

\begin{figure}[thp]
\centering
\includegraphics[width=0.5\textwidth,page=1]{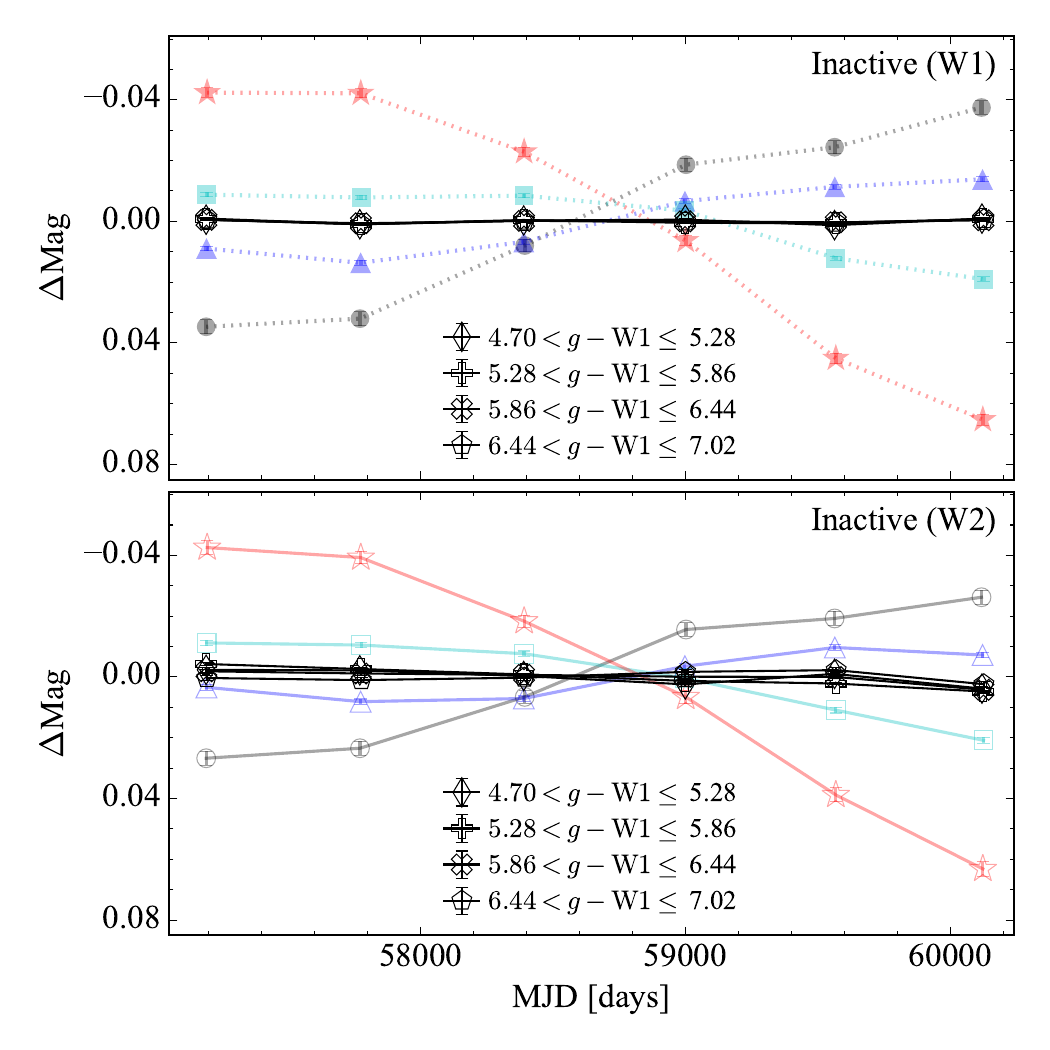}
\caption{Average composite light curves in the W1 (top) and W2 (bottom) bands for inactive galaxies (black open symbols) grouped by $g-$W1 color. The average composite light curves of type 1 AGNs from Figure 6 are overplotted in transparent colors for comparison. 
}
\end{figure}
\noindent including \lbol, \mbh, and \redd. To assess the potential impact of \citet{eddington_1913} bias arising from possible flux boost of faint sources, we applied a magnitude limit of $g<18.5$, which corresponds to $\sim2$ mag brighter than the $5\,\sigma$ detection limit for ZTF, and W1 $<14.8$ and W2 $<13.2$, which are $\sim 1$ mag brighter than the 90\% completeness limit for NEOWISE. We adopted the brighter limit for the $g$-band, as forced photometry in ZTF can suffer from systematic photometric uncertainties for low-S/N sources \cite[e.g.,][]{lang_2016}. We confirmed that the SF asymmetries are still undetected after applying the magnitude limits. Since each subsample in this analysis contains a limited number of targets (mostly fewer than 200 objects), the results should be used solely for verification purposes.
 
Interestingly, however, the SF asymmetry in the W1 and W2 bands becomes prominent when type 1 AGNs are categorized by their optical-to-MIR color. As shown in Figure 4, MIR SFs exhibit positive asymmetry for AGNs with bluer $g-$W1 colors, while negative asymmetry is observed for those with redder $g-$W1 colors. A similar trend was also found in type 2 AGNs, although this result should be interpreted with caution due to the limited number ($<200$) of sources in some subsamples. The degree of SF asymmetry is particularly pronounced at longer timescales, above a few hundred days. A comparable trend was observed in the $g$-band SF for type 1 AGNs, although the asymmetry is significantly weaker than that observed in the MIR SFs. We note that the color dependence of SF asymmetry is evident when subsamples are divided by the optical-to-MIR colors ($g-$W1, $g-$W2, $r-$W1, and $r-$W2), while it is absent or significantly weaker for other colors (e.g., $g-r$ and W1$-$W2). To minimize Eddington bias, we re-examined the SF asymmetry by applying magnitude cuts ($g<18.5$, W1 $<14.8$, and W2 $<13.2$) and found that its dependence on $g-$W1 color in the MIR remains unchanged. Additionally, we confirmed that the dependence on the $g-$W1 color persists even when using the host-subtracted $g-$W1 color (Fig. 5). In contrast, the SF asymmetry in the optical substantially declines when magnitude cuts are applied, indicating that the asymmetry in the $g$-band SF is likely driven by Eddington bias. We discuss the possible origins of MIR SF asymmetry and its dependence on optical-to-MIR color in the following section.

\begin{figure*}[htp!]
\centering
\includegraphics[width=0.9\textwidth]{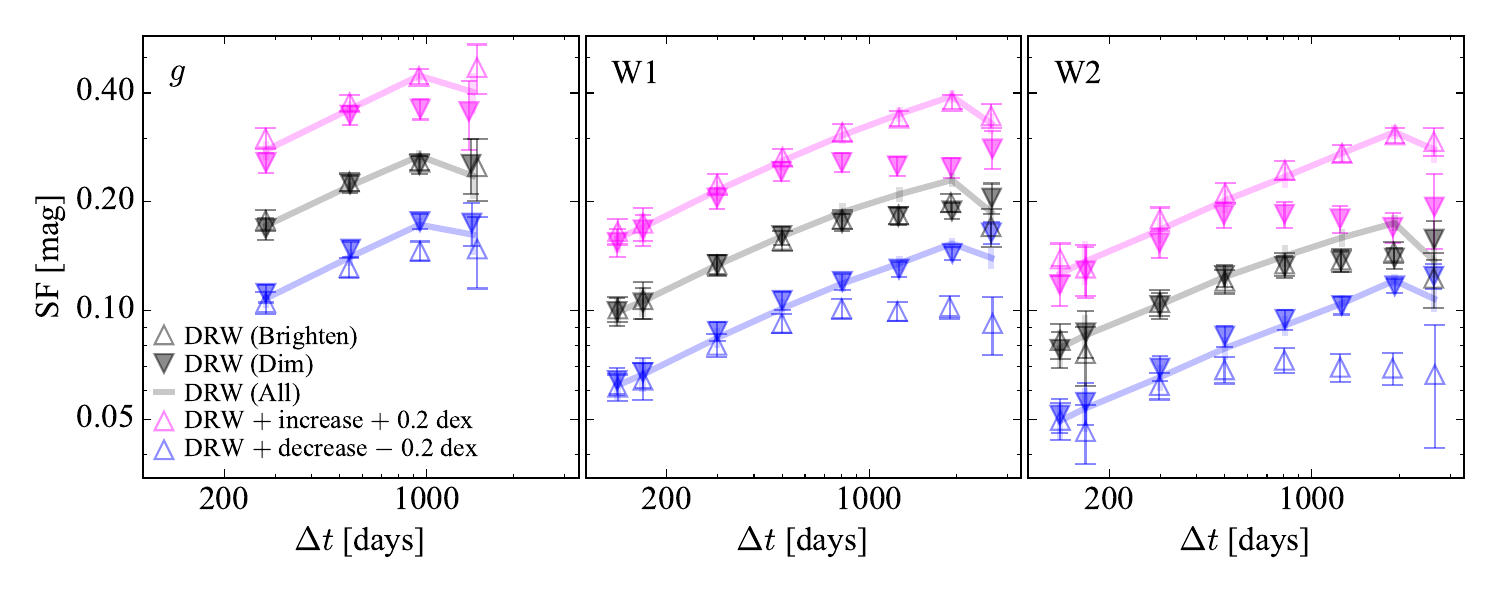}
\caption{Ensemble SFs computed from 200 mock light curves for the $g$ (left), W1 (middle), and W2 (right) bands. Upward and downward triangles denote the ensemble SFs for the brightening and dimming phases, respectively, and the solid line denotes the ensemble SFs for all phases. Black triangles represent the ensemble SFs following a DRW model, while magenta and blue triangles represent those with global increasing and decreasing trends in brightness, respectively. The SF value was shifted arbitrarily to improve the clarity of the visualization. 
}
\end{figure*}

\section{Discussion}

\subsection{What causes the asymmetry in the MIR SF?}

\subsubsection{Average composite light curve}

To investigate the nature of the asymmetry in the SF, we constructed the average composite light curve of each subsample divided by $g-$W1 color. For consistency, we used only targets with more than 18 epochs for NEOWISE and 5 epochs for ZTF, as in the SF calculation. We subtracted the mean magnitude from the light curve of each object to normalize across objects. The composite light curve was then estimated by taking the mean values within each observing time bin defined in the observed frame. The calculated composite light curves are shown in Figure 6. Notably, AGNs with bluer $g-$W1 colors, which exhibit positive asymmetry, tend to show a gradual increase in MIR brightness, while those with redder $g-$W1 colors, showing negative asymmetry, tend to marginally fade over time. This suggests that SF asymmetry is driven by a long-term global trend. The largest magnitude difference ($\sim 0.11$ mag in the W1 and W2 bands) was observed in the type 1 AGN subsample with the reddest $g-$W1 color ($5.4<g-{\rm W1}\leq 6.0$). Note that these trends remained the same even when the time bin was calculated in the rest-frame or when the median magnitude was used instead of the mean value. The global brightening and dimming are unlikely to be artifacts, as they are not found in the average composite light curves of inactive galaxy subsamples, which are computed in the same manner as for the AGNs (Fig. 7).

\subsubsection{Simulation}

To further assess if the temporal asymmetry arises from the global trends or potential systematics of our datasets, we performed a series of simulations replicating the observational conditions. To this end, we generated 200 mock light curves, which we adopt as the minimum number to ensure stable SF asymmetry analysis, assuming, for simplicity, a DRW model with a damping timescale of 660 days. To reflect the larger variability at shorter wavelengths, the SF at $\Delta t \to \infty$ was set to 0.25, 0.2, and 0.15 for the $g$, W1, and W2 bands, respectively. Two sets were constructed: (1) with an imposed global brightening or dimming trends of $\sim 0.11$ mag over the observing duration, and (2) without such a modification (Fig. 8). The value of 0.11 mag was chosen to match the most significant magnitude difference observed in the average composite light curves of type 1 AGNs, as discussed in Section~5.1.1. Finally, to mimic realistic observing conditions, the mock datasets were sampled with the actual cadences, baselines, and photometric uncertainties of the observed data. We then computed the ensemble SF$_+$ and SF$_-$ from the mock light curves. 

As expected, the ensemble SFs computed from light curves without global trends show no difference between the brightening and dimming phases. Owing to the limitations of the baseline and observing cadence, the SF appears to be reliable on timescales of $\Delta t \lesssim 1000$ days for ZTF and $\Delta t \lesssim 2000$ days for NEOWISE, which defines the range used for the SF asymmetry analysis presented in Section 4. The temporal asymmetry was successfully reproduced in the set with injected global brightening and dimming, while the set without such trends recovered the original SFs with negligible systematics. These results suggest that the observed asymmetry is unlikely to arise from observational systematics and is more plausibly driven by intrinsic gradual brightness variations in the sample.

\subsection{Physical origin of the relation between the global trend and optical-to-MIR color}

The analyses of the SF and composite light curve revealed that AGNs with bluer optical-to-MIR colors tend to brighten in the MIR, while redder AGNs tend to dim. If the SF asymmetry in the MIR is the direct reflection of that in the emission from the AD, the timescale at which the asymmetry begins to occur in the MIR should be longer than that in the ultraviolet-optical continuum due to the geometric effect \cite[e.g.,][]{li_2023, kim_2024}. Additionally, the degree of asymmetry in the optical should be larger than that in the MIR. However, the lack of asymmetry in the $g$-band SF at timescales where the MIR SF exhibits significant asymmetry implies that temporal asymmetry in the MIR continuum is unlikely to originate from temporal asymmetry in the AD emission. 

The optical-to-MIR color can serve as an indicator of the relative contribution of hot dust emission compared to AD and host galaxy emission, as the optical continuum originates from the AD and host galaxy, while the MIR continuum primarily originates from the hot dust component of the torus \cite[e.g.,][]{jun_2013,son_2023}. As our analysis used sources with low host galaxy fractions, the optical-to-MIR color may trace the hot dust contribution relative to the AD. This interpretation is supported by the fact that the color dependence of SF asymmetry persists even when using the host-subtracted $g-$W1 color for type 1 AGNs (middle panels of Fig. 5). Therefore, the dependence of temporal asymmetry on optical-to-MIR color is likely to originate from the correlation between the color and the hot dust contribution. 

The optical-to-MIR color dependence of temporal asymmetry may imply that the torus is a self-regulating system. AGNs with bluer optical-to-MIR colors may indicate a lower contribution (or covering factor) from hot dust in the torus. In such cases, the amount of hot dust increases with time, as suggested by the global variation trend in the MIR, and vice versa. If so, the torus appears to have a tendency to revert to its original state, at least on a timescale of years.

These findings may be consistent with the wind-driven torus formation scenario, in which the wind generated from the AD pushes the hot gas outward, leading to the formation of dense clouds through a cooling process \cite[e.g.,][]{elitzur_2006, wada_2012}. In this framework, the brightness between the AD and the torus may be balanced because the outflow strength is proportional to the AGN brightness, thereby maintaining a constant optical-to-MIR color. However, in our observations, the asymmetry in the optical continuum is less pronounced than that in the MIR continuum. This mismatch between the optical and MIR may suggest that an additional mechanism is required to enhance the asymmetry in the MIR. One possibility is that the response of the torus is somewhat delayed, probably due to the dynamical timescale of the outflows (\citealt{elvis_2002}). Although the above explanation remains speculative, it nevertheless suggests a causal connection between the AD and the torus. Therefore, our findings may disfavor the ex-situ torus formation scenario, in which the dust in the torus originates from the accretion of gas clouds from the outer regions of the central AGN \cite[e.g.,][]{krolik_1988}.

Additionally, the enhanced temporal asymmetry in the MIR compared to AD variability may reflect that the timescales of destruction and formation of hot dust in response to the AD continuum are substantial. Although the heating and cooling timescales of the hot dust (on the order of $10^{-6}$ seconds and less than a few seconds, respectively; \citealt{nenkova_2008b, vanvelzen_2016}) are negligible compared to the typical survey cadence, the destruction and formation of the hot dust have been reported to occur on timescales of years \citep{oknyanskij_2006, koshida_2009, kishimoto_2013}. Compared to earlier studies, \cite{kishimoto_2013} presented the most robust constraint on the timescale of hot-dust destruction and formation ($\sim 6$ years), by showing that the interferometric radius of the hot dust component exhibits the strongest correlation with the optical flux averaged over the past 6 years. Our findings suggest that AGNs with higher AD-to-torus flux ratios (e.g., bluer optical-to-MIR colors) may be undergoing hot-dust formation over a timescale of years, whereas those with lower AD-to-torus flux ratios (e.g., redder optical-to-MIR colors) may be undergoing hot-dust destruction. Global brightening and dimming are observed throughout the composite light curves, which span $5.7-7.0$ years in the rest-frame. However, this observed timescale is difficult to interpret directly in terms of formation and destruction timescales, because the baseline may not be sufficient to cover the entire brightening and dimming phase, resulting in a lower bound. Although the relevant timescales and the physical origin of global trends in the MIR remain unclear, our findings offer critical insights into the evolution of the dusty torus.

\subsection{Potential origin of symmetric variation in the optical light curve}
Notably, the symmetry is evident in the $g$-band light curves (Fig. 3), which can potentially provide useful insights on the physical origin of the variability in the accretion disk. While the symmetry can naturally arise from microlensing \citep{hawkins_2002}, it could also be caused by disk instabilities or starbursts, where the degree of asymmetry is sensitive to model parameters (e.g., the ratio of diffusion mass to inflow mass and the supernova rate; \citealt{kawaguchi_1998}). Additionally, the short baseline of the light curve may simply prevent the identification of any true underlying asymmetry. Therefore, we cannot draw a firm conclusion about the physical origin of the variability based solely on the current dataset.

Interestingly, MIR variability also exhibits symmetry across the entire sample of AGNs (Fig. 3). However, the strong dependence of the MIR variability on the optical-to-MIR color indicates that this symmetry is unlikely to be a simple consequence of the symmetry observed in the optical light curve. Moreover, microlensing may not be a direct cause of the MIR symmetry either, as the physical size of the torus is too large compared to the Einstein radius of the potential lens. The observed symmetry in the MIR SF may therefore be a coincidence, possibly arising from a statistical balance between hot-dust-rich and hot-dust-poor AGNs.

\section{Summary}

To explore temporal asymmetry in the torus variability for the first time, we analyzed the MIR and optical continuum variability of AGNs using NEOWISE (W1 and W2 bands) and ZTF ($g$-band) light curves. We compared ensemble SFs between brightening and dimming phases for subsamples divided by various properties. The following conclusions are drawn from our analysis.

\begin{itemize}
    \item SF asymmetry in the MIR variability is most evident when AGNs are divided into subsamples based on their optical-to-MIR color. AGNs with bluer optical-to-MIR colors show positive asymmetry (i.e., ${\rm SF_+>SF_-}$), whereas AGNs with redder colors show negative asymmetry (i.e., ${\rm SF_->SF_+}$).

    \item For type 1 AGNs, where the $g$-band continuum is dominated by the AD, no significant asymmetry was observed in the $g$-band SF. This suggests that the asymmetry in the MIR SF is unlikely to be directly driven by variability in the AD, but instead originates from physical processes in the torus.

    \item The composite light curves of the subgroups indicate that SF asymmetry in the MIR is primarily driven by global brightening and dimming phases. AGNs with bluer optical-to-MIR colors tend to exhibit gradually rising light curves, whereas those with redder colors show declining light curves.
    
    \item As the optical-to-MIR color is a useful tracer of the hot dust contribution relative to the AD, the dependence of SF asymmetry on color may indicate that the hot dust component in AGNs acts as a self-regulating system: AGNs rich in hot dust tend to lose hot dust, while those poor in hot dust tend to accumulate more hot dust.

\end{itemize}

\begin{acknowledgements}
We are grateful to the anonymous referee for the constructive comments that have significantly improved the manuscript. LCH was supported by the National Science Foundation of China (12233001) and the China Manned Space Program (CMS-CSST-2025-A09). This work was supported by the National Research Foundation of Korea (NRF) grant funded by the Korean government (MSIT) (Nos. RS-2024-00347548 and RS-2025-16066624).  
\end{acknowledgements}

\bibliography{torus}

\end{document}